\documentclass[twocolumn,amsmath,amssymb,floatfix,prl,showpacs]{revtex4}
\usepackage{bm}
\newcommand{\bomega}{\bm\omega}
\begin{document}
\title{%
Oscillations of 2DEG thermoelectric coefficients in magnetic field under
microwave irradiation%
}
\author{A. E. Patrakov}
\author{I. I. Lyapilin}
\email{Lyapilin@imp.uran.ru}
\affiliation{Institute of Metal Physics, UD of RAS, Yekaterinburg, Russia}
\begin{abstract}
It is known that under microwave irradiation, in 2D electron systems with high
filling factors oscillations of longitudinal magnetoresistance
appear in the range of magnetic fields where ordinary SdH
oscillations are suppressed.
In the present paper we propose a simple
quasiclassical model of these new
oscillations based on
the Boltzmann kinetic equation. Our model also predicts similar oscillations
in diffusion component of thermoelectric coefficients,
which should be observable at low temperatures.
\end{abstract}
\maketitle

\section{Introduction}
The interest to theoretical investigations of non-linear transport
phenomena in two-dimensional electron systems (2DES) has grown
substantially due to new experimental results obtained in very
clean 2DES samples. It has been discovered independently by two
experimental groups \cite{Zudov, Mani} that the resistance of
two-dimensional high-mobility electron gas in $GaAs/AlGaAs$
heterostructures reveals a series of new features in its
dependence on magnetic field $H$ , temperature $T$, radiation power, etc.
Under microwave irradiation, in 2DES with high filling factors
oscillations of longitudinal magnetoresistance appear in the range
of magnetic fields where ordinary SdH oscillations are suppressed.
Under high-intensity irradiation, the minima of these oscillations
become zero-resistance states. Unlike the SdH oscillations, which depend
upon the ratio of the chemical potential $\zeta$ to the cyclotron frequency
$\omega_c$,
these oscillations caused by irradiation depend upon the ratio of
the radiation frequency $\omega$ to the cyclotron one.
A series of theoretical papers
\cite{ShiXie, Durst, Ryzhii, Mikhailov} has been submitted that aim to
explain these oscillations.

Let's mention important conditions of the experiments
discussed above: the effect is seen at
$$\hbar/\tau\ll T \simeq \hbar\omega_{\rm c} \leq \hbar\omega \ll
\zeta,$$
where $\tau$ is the transport relaxation time.
It follows that the effect has a quasiclassical nature.
Since the oscillations express themselves in the classical range of magnetic
fields, it is reasonable to consider this effect using a model based on the
Boltzmann kinetic equation. The Boltzmann kinetic equation is also capable
of describing thermoelectric phenomena, and our model predicts oscillations
of thermopower and Nernst --- Ettingshausen coefficient controlled by
the $\omega/\omega_{\rm c}$ ratio. They should be observable at low
temperatures.

However, in this paper only contribution of diffusion processes into
thermoelectric coefficients is considered. At liquid helium temperatures,
scattering of electrons upon nonequilibrium phonons and their mutual drag is
also important. The contribution of processes involving phonons to
thermoelectric coefficients in the presence of microwave radiation will be
considered elsewhere.

\section{Model}
Taking into account that the electrons are driven from the
equilibrium both by the temperature gradient,
the external dc electric field and the electric
field of radiation, we have the following kinetic equation for electrons in
the 2DES:
\begin{multline}
\frac{\bm p}{m}\frac{\partial f(\bm r, \bm p)}{\partial \bm r}+
\left.\frac{\partial f(\bm r, \bm p)}{\partial t}\right|_{E_{\rm ac}} -e \bm
E_{\rm dc} \frac{\partial f(\bm r, \bm p)}{\partial \bm p}-\\
\frac{e}{m c}\left[
\bm p \times  \bm H \right] \frac{\partial f(\bm r, \bm p)}{\partial \bm
p} = I_{\rm st}[f(\bm r, \bm p)].
\end{multline}
Here $E_{\rm dc}$ is the strength of the external dc field.
The right hand side is the collision integral. We assume that the main
electron scattering mechanism is the elastic scattering on impurities:
\begin{equation}
I_{st}[f(\bm r, \bm p)] = - \frac{f(\bm r, \bm p)-f_0(\bm r, \bm p)}{\tau},
\end{equation}
where $f_0(\bm r, \bm p)$ is the equilibrium distribution function,
$\tau$ is the relaxation time.
$\left.\frac{\partial f(\bm r, \bm
p)}{\partial t}\right|_{E_{\rm ac}}$
is the rate of distribution function change due to transitions between Landau
levels caused by microwave radiation:
\begin{equation}
\left.\frac{\partial f(\bm r, \bm p)}{\partial t}\right|_{E_{\rm ac}}=
\sum_{\bm p'} w_{\bm p \bm p'} (f(\bm r, \bm p') - f(\bm r, \bm p)),
\end{equation}
where $w_{\bm p \bm p'}$ is the probability of the electron transition
from the state with the kinetic momentum $\bm p$ to the state with the
kinetic momentum $\bm p'$ in a unit of time due to microwave radiation.

Up to the first order on thermodynamic forces, the rate of distribution
function change resulting from diffusion due to the temperature gradient is:
\begin{equation}
\frac{\bm p}{m}\frac{\partial f(\bm r, \bm p)}{\partial \bm r} =
-\frac{\bm p}{m} T \frac{\partial f_0(T(\bm r), \varepsilon(\bm p))}
{\partial \varepsilon(\bm p)}
\frac{\partial}{\partial \bm r}\left(
\frac{\varepsilon(\bm p) - \zeta}{T(\bm r)}
\right).
\end{equation}

The non-equilibrium distribution function can be represented in the following
form:
\begin{equation}\label{fdist}
f(\bm r, \bm p) =
f_0(T(\bm r), \varepsilon(\bm p)) + \bm p \bm g(\varepsilon(\bm p)),
\end{equation}
where $f_0(T(\bm r), \varepsilon(\bm p))$ is the equilibrium
distribution function
and $\bm g(\varepsilon(\bm p))$ is an unknown function that depends only on
electron energy.

The main effect of the ac electric field amounts to changes in electron
energy (and not its momentum).

\section{Distribution function and current density}
Taking all simplifications mentioned above into
account and linearizing upon thermodynamic forces,
we can express the kinetic equation in the following form:
\begin{multline}
\label{tosolve} \sum_{\bm p'} w_{\bm p \bm p'} (\bm
g(\varepsilon') - \bm g(\varepsilon)) -\frac{1}{m} \mathcal F_{\rm dc}
\frac{\partial f_0(T, \varepsilon)}{\partial \varepsilon} -\\
\frac{e}{m c}\left[ \bm H \times \bm g(\varepsilon) \right] +\frac{\bm
g(\varepsilon)}{\tau}  = 0,
\end{multline}
where we denoted
\begin{equation}
\mathcal F_{\rm dc} = e \bm E_{dc} + \frac{\varepsilon - \zeta}{T} \bm \nabla T.
\end{equation}

We shall search for a solution of \eqref{tosolve} in the form of a power
series on $w_{\bm p \bm p'}$. Up to the linear terms, we have:
\begin{multline}
\label{soldg}
\bm g(\varepsilon) = \bm g_0(\varepsilon) -
\frac{\tau^2}{(1+\omega_{\rm c}^2 \tau^2)^2}
\frac{1}{m}\times\\
((1-\omega_{\rm c}^2\tau^2) \mathcal F_{\rm dc} +
2 \tau[\bomega_{\rm c} \times \mathcal F_{\rm dc}])\times\\
\sum_\pm \rho(\varepsilon\pm\hbar\omega) w_\pm
\left(\frac{\partial f_0(\varepsilon\pm\hbar\omega)}{\partial
\varepsilon} -\frac{\partial f_0(\varepsilon)}{\partial
\varepsilon}\right),
\end{multline}
where
\begin{equation}
\label{iter1solved} \bm
g_0=\frac{\tau}{1+\omega_{\rm c}^2\tau^2}\frac{1}{m} (\mathcal F_{\rm dc}
+ \tau[\bomega_{\rm c} \times \mathcal F_{\rm dc}]) \frac{\partial
f_0(\varepsilon)}{\partial \varepsilon}.
\end{equation}
Here $\bomega_{\rm c}=(e \bm H)/(m c)$ and it has been taken into account
that $\bm \nabla T, \ \bm E_{\rm dc} \perp \bm H$.
The probability of transition with
absorption or emission of a photon is
$w_\pm \propto E^2_{\rm ac} \ell^2 \left| \langle n \pm 1 | x | n
\rangle \right| ^2 \propto E^2_{\rm ac} n / \omega_{\rm c} \propto E^2_{\rm ac}
\varepsilon_n / \omega_{\rm c}^2$, where $E_{\rm ac}$ is the amplitude of
the ac electric field with the frequency of
$\omega$, $\rho(\varepsilon)$ is the density of states, and $\ell=\sqrt{\hbar
/ (m \omega_{\rm c})}$ is the magnetic length.

Using the correction to the distribution function calculated above
we find the current density caused by radiation:
\begin{equation}\label{current}
\bm j = - \frac{2 e}{(2 \pi \hbar)^2 \rho_0}
\int d \bm p \ \bm p (\bm p \bm g(\varepsilon(\bm p))),
\end{equation}
where $\rho_0=m/(2\pi\hbar^2)$ is the density of
states without the magnetic field for one spin direction.

The current can originate from the dc electric field or due to the
temperature gradient. In linear theory, the relation of current density with
thermodynamic forces is:
\begin{equation}
j_k = \sigma_{kl} E_{{\rm dc}, l} - \beta_{kl} \nabla_l T
\end{equation}

Inserting \eqref{soldg} into \eqref{current} and putting $\bm \nabla T = 0$,
we obtain for diagonal components of the conductivity tensor:
\begin{multline}
\sigma_{\rm xx}^{\rm ph} =
- \frac{2 e^2}{m} \frac{\tau^2 (\omega_{\rm c}^2\tau^2 -
1)}{(1+\omega_{\rm c}^2 \tau^2)^2}
 \sum_\pm
\int_0^\infty d\varepsilon \ \varepsilon \times\\
w_\pm
\rho(\varepsilon) \rho(\varepsilon\pm\hbar\omega)
\left(\frac{\partial f_0(\varepsilon\pm\hbar\omega)}{\partial
\varepsilon} -\frac{\partial f_0(\varepsilon)}{\partial
\varepsilon}\right)
\end{multline}

Taking the energy integral under the assumption of strong degeneracy of
electrons and using an explicit expression for the
density of states
\cite{Ando}
\begin{eqnarray}
\rho(\varepsilon)=\rho_0
(1-\delta\cos(\frac{2\pi\varepsilon}{\hbar\omega_{\rm c}})),
\nonumber\\
\delta=2 e^{-\pi/(\omega_{\rm c}\tau_{\rm f})} \ll 1
\end{eqnarray}
($\tau_{\rm f}$ is a single-particle lifetime without magnetic field)
we obtain the following expression for longitudinal photoconductivity:
\begin{equation}\label{full}
\sigma_{\rm xx}^{\rm ph}-\sigma_{\rm xx}^{\rm ph,\ SdH} \propto
\frac{\omega^2\tau^2 (\omega_{\rm c}^2\tau^2-1)}{\omega_{\rm c}^2
(1+\omega_{\rm c}^2\tau^2)^2} \delta^2 \cos\frac{2\pi\omega}{\omega_{\rm c}}
\end{equation}

For simplicity, we denote all factors in
$\sigma_{\rm xx}^{\rm ph}-\sigma_{\rm xx}^{\rm ph,\ SdH}$ except
$(\omega_{\rm c}^2\tau^2 -1) \cos (2\pi\omega/\omega_{\rm c})$ with the
letter $A$ ($A$ is proportional to the microwave power):
\begin{equation}
\sigma_{\rm xx}^{\rm ph}-\sigma_{\rm xx}^{\rm ph,\ SdH} =
A (\omega_c^2\tau^2 -1) \cos \frac{2\pi\omega}{\omega_{\rm c}}
\end{equation}

Performing similar calculations, one obtains for Hall photoconductivity:
\begin{equation}
\sigma_{\rm xy}^{\rm ph}-\sigma_{\rm xy}^{\rm ph,\ SdH} =
2 A \omega_{\rm c} \tau \cos \frac{2\pi\omega}{\omega_{\rm c}}
\end{equation}
with the same factor $A$ as above.

Then we calculate the change of $\beta$ tensor components due to microwave
irradiation. To do this, we insert \eqref{soldg} into \eqref{current} and
put $\bm E_{\rm dc} = 0$. The result, under the same assumptions, is:
\begin{equation}
\beta_{\rm xx}^{\rm ph}=\frac{2 A \zeta (1-\omega_{\rm c}^2\tau^2)}{e T}
\cos\left(\frac{2\pi\omega}{\omega_{\rm c}}\right)
\end{equation}
\begin{equation}
\beta_{\rm xy}^{\rm ph}=-\frac{4 A \zeta \omega_{\rm c}\tau}{e T}
\cos\left(\frac{2\pi\omega}{\omega_{\rm c}}\right)
\end{equation}

In thermoelectric experiments,
one usually does not allow the current to flow in the sample, applies the
temperature gradient, and measures the voltage between the opposite edges of
the sample. To find the electric field that develops due to the temperature
gradient in such conditions, one should solve the following system of
equations:
\begin{multline}
j_{\rm x} =\sigma_{\rm xx}^0 E_{\rm dc,\ x}
+\sigma_{\rm xx}^{\rm ph} E_{\rm dc,\ x}
+\sigma_{\rm xy}^0 E_{\rm dc,\ y}
+\\
\sigma_{\rm xy}^{\rm ph} E_{\rm dc,\ y}
-\beta_{\rm xx}^0 \nabla_{\rm x} T
-\beta_{\rm xx}^{\rm ph} \nabla_{\rm x} T
=0
\end{multline}
\begin{multline}
j_{\rm y}=\sigma_{\rm xx}^0 E_{\rm dc,\ y}
+\sigma_{\rm xx}^{\rm ph} E_{\rm dc,\ y}
-\sigma_{\rm xy}^0 E_{\rm dc,\ x}
-\\
\sigma_{\rm xy}^{\rm ph} E_{\rm dc,\ x}
+\beta_{\rm xy}^0 \nabla_{\rm x} T
+\beta_{\rm xy}^{\rm ph} \nabla_{\rm x} T
=0,
\end{multline}
where $\sigma^0$ and $\beta^0$ denote the corresponding coefficients without
microwave irradiation.

Solving this linear system and taking into account the following relations:
$$\sigma_{\rm xy}^0 = - \omega_{\rm c} \tau \sigma_{\rm xx}^0,$$
$$\beta_{\rm xy}^0 = - \omega_{\rm c} \tau \beta_{\rm xx}^0,$$
and also $\zeta \gg T$, we find:
\begin{equation}\label{tp}
E_{\rm dc,\ x} = \left( \frac{\beta_{\rm xx}^0}{\sigma_{\rm xx}^0} +
\frac{2 \zeta}{e T \sigma_{\rm xx}^0}
A \cos \left( \frac{2 \pi\omega}{\omega_{\rm c}} \right)
+O(A^2) \right) \nabla_{\rm x} T
\end{equation}
\begin{equation}\label{ne}
E_{\rm dc,\ y} = \left( \frac{2 \zeta \omega_{\rm c} \tau}
{e T \sigma_{\rm xx}^0}
A \cos \left( \frac{2 \pi\omega}{\omega_{\rm c}} \right)
+O(A^2) \right) \nabla_{\rm x} T
\end{equation}

\section{Conclusion}
One can see from Eqs. \eqref{tp}, \eqref{ne} that the diffusion components
of thermoelectric coefficients (namely, thermopower and the
Nernst --- Ettingshausen coefficient) in the presence of microwave irradiation
should oscillate with the same period in $1/B$ as magnetoresistance does.
This effect should be better observed at low temperatures, when the
thermopower is low by itself, and the correction is large due to the
$1/T$ factor, unusual for diffusion processes.

\newcommand{\PRL}{{\it Phys. Rev. Lett.}}

\end{document}